\documentclass{ws-ijmpd}
\usepackage[super,compress]{cite}
\usepackage{pstricks}
\begin{document}


%
\catchline{}{}{}{}{}
%

\title{REGULAR BLACK HOLES FROM SEMI-CLASSICAL DOWN TO PLANCKIAN SIZE}
\author{EURO SPALLUCCI}

\address{Department of Physics, University of Trieste, Strada Costiera 11\\
Trieste, Italy 34151,
Italy, and\\
I.N.F.N. Sezione di Trieste, Strada Costiera 11\\
Trieste, Italy 34151,
Italy
\\
euro@ts.infn.it}

\author{ANAIS SMAILAGIC}

\address{ I.N.F.N. Sezione di Trieste, Strada Costiera 11\\
Trieste, Italy 34151,
Italy\\
anais@ts.infn.it}

\maketitle

\begin{history}
\received{Day Month Year}
\revised{Day Month Year}
\end{history}

\begin{abstract}
In this paper we review various models of curvature singularity free black holes. In the first part of the review we describe semi-classical
solutions of the Einstein equations which, however, contains a ``~quantum~'' input through the matter source. We start by reviewing the
early model by Bardeen where the metric is regularized by-hand through a short-distance cut-off, which is justified in terms
of non-linear electro-dynamical effects. {  This  toy-model is useful to point-out the common
features shared by all  regular semi-classical black holes.} Then, we solve Einstein equations with a Gaussian source encoding the quantum
spread of an elementary particle. We  identify, the a priori arbitrary,  Gaussian width  with the Compton wavelength of the quantum particle.
This Compton-Gauss model leads to the estimate of a terminal density that a gravitationally collapsed object can achieve.  We identify
this density to be the Planck density,  and reformulate the Gaussian model assuming this as its peak density. All these models, are physically
reliable as long as the black hole mass is big enough with respect to the Planck mass. In the truly Planckian regime, the semi-classical
approximation breaks down. In this case, a fully quantum black hole description is needed. In the last part of this paper, we propose
a non-geometrical quantum model of Planckian black holes implementing the Holographic Principle and realizing the ``~classicalization~''  
scenario recently introduced by Dvali and collaborators. The classical relation between the mass and radius of the black hole emerges only in 
the classical limit, far away from the Planck scale.

\end{abstract}


\ccode{  PACS numbers: 04.70.Dy,04.70.Bw,04.60.Bc}

\tableofcontents

\section{Introduction}	

The attribution of the copyright for the term ``black hole'' describing the end-point of massive astrophysical gravitationally collapsed bodies
is still an open issue. It seems the term was used for the
first time  in December 1963 at the First Texas Symposium by... someone!\\ 
However, the idea of ``hidden stars'' dates back to Laplace \footnote{``\textit{There exist in the heavens therefore dark bodies, as large
and perhaps as numerous as the stars themselves. Rays from a luminous star having the same density as the Earth and a diameter
250 times that of the Sun would not reach us because of its gravitational attraction; it is therefore possible that the largest
luminous bodies in the Universe may be invisible for this reason}''\cite{laplace}.} 
and Michell \footnote{``\textit{If the semi diameter of a sphere of the same density of the Sun were to exceed that of the Sun
in proportion of 500 to 1, a body falling from an infinite height toward it, would have acquired at its surface a greater velocity than
light, and consequently, supposing light to be attracted by the same force in proportion to its vis inertiae, with other bodies, 
all light emitted  from such a body would be made to return towards it, by its own proper gravity}''\cite{michell}.}. 
They argued that a uniform density star, massive enough, exerts a Newtonian gravitational pull sufficiently strong  to prevent even 
light rays from leaving its  surface. The matter density of a hidden star, of mass $M$, is evaluated to be

\begin{equation}
 \rho =\frac{3c^6}{32\pi G_N^3}\frac{1}{M^2}
\label{rhomax}
\end{equation}

This is a Newtonian result following from the requirement that the escape velocity cannot exceed the speed of light. 
The resulting radius  is  $R=2MG_N/c^2$, what is today known as the Schwarzschild radius. 
Thus, even in this classical version the  ``~black hole progenitor~'' results to be a
 finite density, massive, objects of the size as the one obtained from General Relativity (GR). \\
Furthermore, (GR) modified the Newtonian picture by showing  that 
 classical matter cannot be stationary inside a black hole (BH). Thus, collapsing matter  unavoidably ends-up 
into an infinite density ``~singularity~''. In this picture general relativistic BHs are depicted as \emph{``~empty~''} objects with the
total mass concentrated in a zero volume space region. On the other hand, singularities are the loci where the predictability of the 
theory breaks down and the very definition of the  source of the field becomes ill defined... to use an euphemism. 
To avoid unpredictability of the theory, it has been conjectured by R.Penrose in the late sixties that,  a yet unknown physical principle,
 colloquially named   the ``~Cosmic Censorship~'', would prevent the physical realization of any ``~naked~'' curvature singularity exposed 
to asymptotic observers. Until today it has remained a  ``~conjecture~''.\\

Since the advent of the  solution  by Schwarzschild of the Einstein equations, BHs have exercised a fascinating appeal of a mysterious 
cosmic objects where the very concepts of space and time, as we know them, loose meaning. Beside being the favorite theme in a long 
sequence of science fiction novels and movies, a parallel and equally ``\emph{mythological}'' line of thought has taken ground among physicists.
 This is probably due to an excessive eagerness  to find mathematical solutions of Einstein equations without an \emph{a priori} clear physical 
input. The interested reader can find a clear  example of this approach both in the original paper by Kerr \cite{Kerr:1963ud}, in the case
of a rotating BH, as well as, in the further ``~clarification'' by S.Chandrasekhar \cite{Chandra}.\\
This purely mathematical approach has, on its own, led to  the following misconceptions:
\begin{romanlist}[(iiii]
 \item BHs are  solutions of the Einstein equations without a matter source, i.e. they are ``vacuum'' solutions;
 \item the sole existence of the  BH horizon implies the presence of a curvature singularity in its interior.
 \item BHs are completely``empty''.
 \item Even at the Planck energy,  BHs are still described in a classical geometric way.
\end{romanlist}

(i) It is mathematically legitimate to solve the field equations outside the source, i.e. in ``~vacuum~''. However, it is not correct 
to forget about the existence of any source,  and claim that the energy and momentum distribution is  vanishing everywhere!  
Take a look at whatever textbook in General Relativity and check that
the Schwartzschild metric is always introduced by setting $T^{\mu\nu}=0$ in the r.h.s. of the field equations. In the case of a static, 
spherically symmetric, asymptotically flat geometry, one finds a well known line element 

\begin{equation}
 ds^2= -\left(\, 1 -\frac{r_0}{r}\,\right)dt^2 +\left(\, 1 -\frac{r_0}{r}\,\right)^{-1}dr^2 +r^2\left(\, d\theta^2
 + \sin^2\theta d\phi^2\,\right)
\end{equation}

where, $r_0$ is an arbitrary integration constant. A posteriori, $r_0$ is identified with $2M G_N$ in order to recover the Newtonian limit
in the weak field approximation.
But, it is well known that the Newtonian potential is sourced by a spherical mass distribution. Thus, starting with $T^{\mu\nu}=0$, 
one ends-up with a solution with a mass of the source, i.e. a $T^{\mu\nu}\ne 0$, a surprising way to introduce BHs, to say the least! \\
To obtain the Schwarzschild solution, in this way, seems to rely on a series of mathematical manipulations which make mass and curvature
 appear from....  nowhere! {  The correct approach has been discussed in \cite{Balasin:1993fn,Balasin:1993kf,DeBenedictis:2007bm} 
in terms of distributional energy-momentum tensor, but is largely ignored.}

Starting from the physical vacuum,  one expects  Minkowski space-time as the only possible solution. Instead one finds a curved metric, 
which in addition has  a divergent curvature  in $r=0$, where all the mass is concentrated. The conclusion is that the physical source of the 
field is located just in the point where the whole theory looses its meaning.  

(ii) We shall consider in this paper, various types of BH solutions of the Einstein equations, which are curvature singularity free and
correspond to non point-like matter sources.\\

(iii) An immediate consequence of the existence of the singularity, is that the BHs are ``empty'' inside the horizon. {  In fact, 
any object crossing the horizon will ``~disappear~'' inside the singularity  within a proper time lapse of the order of  $10^{-5} M/M_{\odot}$ 
\footnote{ $M_{\odot}$ is the solar mass}, leaving no trace inside the BH.} 
Consequently, there is no stationary mass distribution inside the horizon. Everything ``rains'' towards the singularity. 
 \\
The description  of the BH as a singular object is certainly physically inappropriate, as any physical quantity diverges on the singularity
and the  whole theory looses any predictive power.  \\

(iv)  The problems of describing BHs in a classical geometric way, even at the Planck scale, corresponds to
forcing General Relativity at this energy scale. However, it is widely believed that at this energy,  Quantum Mechanics must be, 
necessarily, taken into account. This issue will be addressed throughout the paper.\\
In General Relativity matter is represented as a  continuous, infinitely compressible, ``~fluid~''.
On the other hand, we know that this picture is inappropriate
 at the quantum level, where Heisenberg and Pauli Principles dictates the behavior of matter building blocks. \\
The Exclusion Principle on its own assures the hydrodynamical stability of white dwarfs and neutron stars. Nevertheless, 
beyond the Oppenheimer-Volkoff limit  self-gravitating masses will continue  to collapse under  their own weight, ending up
ia a BH. \\
In a previous series of papers, we have shown that singular (unphysical) configurations can be avoided
once the idealized  ''~point-like~`` structures are given-up         
\cite{Nicolini:2005vd,Ansoldi:2006vg,Rizzo:2006zb,Nicolini:2008aj,Nicolini:2009gw,Spallucci:2009zz,Smailagic:2010nv,Spallucci:2011rn,Mureika:2011hg},
\cite{Nicolini:2011fy,Nicolini:2012fy,Spallucci:2012xi,Nicolini:2013ega,Spallucci:2014kua,Spallucci:2015jea}. No curvature
singularities appear out of the Einstein field equations in these models. The infinite density limit was avoided due to the presence
of a \emph{minimal} length parameter inspired by different quantum gravity arguments 
\cite{Padmanabhan:1996ap,Fontanini:2005ik,Spallucci:2005bm,Aurilia:2013mca,Kothawala:2014fva,book1}. \\
General Relativity on its own provides a genuine unit of length, time and energy, 
 built out  of the three basic constant: $c$, $G_N$, $\hbar$ \cite{Gibbons:2002iv,Barrow:2014cga}.  

\begin{eqnarray}
 && {  l_{Pl}}=\sqrt{\frac{2\hbar G_N}{c^3}}= 10^{-33}\, cm \ ,\hbox{ length}\label{lp}\\
 && {  t_{Pl}}=\sqrt{\frac{2\hbar G_N}{c^5}}= 10^{-44}\, s \ ,\hbox{ time}\label{tp}\\
 && {  M_{Pl}}=\sqrt{\frac{\hbar c}{2G_N}}= 10^{19}\, GeV = 10^{-5}\, gr \ ,\hbox{ Energy} \label{ep}
\end{eqnarray}

From this vantage point, we propose a physically reliable alternative approach to the ''~singular collapse~`` scenario, 
resulting in an intriguing ''~coexistence~``, and refinement, of both Newtonian and Relativistic model ideas. \\
As far as the density is concerned, the Newtonian result (\ref{rhomax}) indicates the  way how to eliminate curvature singularities.
In other words, every gravitationally collapsing object should end up in a \emph{maximal,finite density} state. 
In terms of the fundamental  units (\ref{lp}), (\ref{tp}), (\ref{ep}) one can define the \emph{Planck Density} as

\begin{equation}
 \rho_{Pl}\equiv \frac{M_{Pl}}{4\pi l_{Pl}^3/3}=\frac{3}{16\pi G_N^2}
\end{equation}

There is a large consensus  that at the Planck scale one has to abandon the usual approach of
point-like particles in a smooth space-time. From this common believe, various different theories emerge, e.g. string theory, loop quantum 
gravity, non-commutative
geometry, etc. However, none of them has succeeded , so far, to give satisfactory answers to the singularity problem,
because, at the end of the day, they rely on classical, though more complicate, smooth geometries to describe ''quantum`` BHs
\footnote{An interesting, alternative, approach, where Planckian BHs are modeled by graviton BECs, with no reference to any (semi/)classical
 background geometry, has been recently proposed in 
\cite{Dvali:2010bf,Dvali:2010jz,Dvali:2011nh,Dvali:2011th,Dvali:2012mx,Dvali:2014ila, Dvali:2011aa,Dvali:2012gb,Dvali:2012rt,Dvali:2012en}
\cite{Casadio:2013hja,Dvali:2013vxa,Dvali:2015aja,Frassino:2016oom,Spallucci:2016ilx} }. \\

In this paper, we shall review various ways to incorporate quantum effects that result in singularity-free BHs. \\
In Section[\ref{proto}], we start from one of the earliest attempts,due to Bardeen, to build a regular BH  and extract from this toy-model
some general features shared by other regular BHs.\\
In Section[\ref{gbh}], we show how a quantum mechanical  matter distribution also leads to exact, singularity-free solutions of the 
Einstein equations. This model contains two free parameters: the total mass $M$ of the source and the  width $a$ of the Gaussian
matter distribution. \\
In Section[\ref{maxdens}] we shall present an alternative formulation of the Gaussian BHs, inspired by the ideas in \cite{Schiller:1996fw},
where it has been argued that $\rho \le \rho_{Pl}$ for any physical object. In this case, the width of the Gaussian distribution is
expressed in terms of $M$ and the resulting geometry is reduced to contain only one free parameter.\\
Finally, in Section[\ref{breath}] we present an original approach to non-geometric BH quantization  where the horizon is allowed
 to  vibrate harmonically at  the Planck scale. A wave equation for the ''\emph{quantum horizon}`` is recovered and solved. 

\section{Proto-regular Black Holes revised}
\label{proto}
In this section we first give a brief review of one of the early attempts to ``~regularize~'' classical 
BHs \footnote{We do not pretend to give an  exhaustive list of all regular solutions, for which we refer to the existing literature 
(see, for example \cite{Ansoldi:2008jw} and references therein).}.\\
In this approach  the strategy was based on a ``~reverse engineering~'' procedure, first one guesses the form of the metric and then
use Einstein equations to recover the corresponding energy-momentum tensor 
\cite{Bardeen,AyonBeato:1998ub,Hayward:2005gi,Frolov:2016pav,Spallucci:2012xi} for
the source.\\
The starting point is the general form of a static, spherically symmetric, asymptotically flat  geometry
\begin{equation}
 ds^2 = -\left(\, 1 - 2G_N\frac{ m(r)}{r}\,\right) \, dt^2 +\left(\, 1 -2G_N \frac{m(r)}{r}\,\right) ^{-1} dr^2 
+ r^2 \left(\, d\theta^2 + r^2\sin^2\theta d\phi^2\,\right) \label{uno}
\end{equation}

which reproduces the Schwartzschild line element for $m(r)\equiv M = const.$. Different models correspond to different choices of
$m(r)$, in \cite{Bardeen} the function was chosen to be
\begin{equation}
 m(r)=\frac{Mr^3}{(\, r^2 + G_N Q^2)^{3/2}}\label{due}
\end{equation}

where $Q$ is an arbitrary regulator parameter which leads to the  de Sitter geometry at short distances, thus removing the curvature singularity.
In the original Bardeen paper the parameter $Q$ was identified with the electric charge based on the hypothesis that a kind of non-linear 
electrostatic repulsion is responsible for the removal of the singularity.\\
Looking at the short-distance behavior of $m(r)$, one finds that 

\begin{equation}
 g_{00}=-g_{rr}^{-1}\to 1 - \frac{2M r^2}{\sqrt{G_N} Q^3}
\label{dslim}
\end{equation}

which is nothing else that a deSitter metric. In this prospective it is natural to relate $Q$ to the  cosmological constant $\Lambda$, i.e. the
vacuum energy density, rather than to the electric charge:

\begin{equation}
 Q^3 =\frac{6M}{\sqrt{G_N}\Lambda}
\end{equation}
The vacuum energy density is characterized by a negative pressure preventing a singular end-state of matter.\\
It will be shown later on that the de Sitter central core is a common characteristic of all regular solutions sourced by a finite, non-vanishing,
 energy density in the origin, $\rho(0)\equiv \rho_0 < \infty $.\\
Another common feature of this kind of regular solutions is the existence of multiple horizons, instead of a single one.
To substantiate this statement, let us look at the horizon equation for the metric (\ref{uno}), (\ref{due}):

\begin{equation}
 M= \frac{1}{2G_N }\frac{\left(\, r_h^2+ Q^2 G_N\,\right)^{3/2}}{r_h^2}\label{mbardeen}
\end{equation}

\begin{figure}[pb]
\centerline{\psfig{file=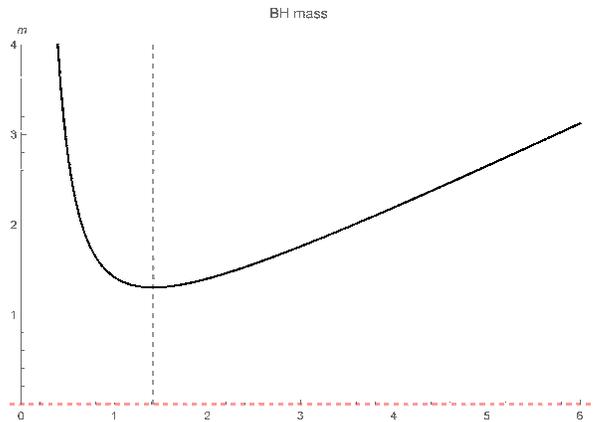,width=8cm}}
\vspace*{8pt}
\caption{Plot of equation (\ref{mbardeen}) showing the existence of an extremal configuration. For $M< M_{extr.}$ there are
no horizons, while for $M> M_{extr.}$ there are two, non-degenerate, horizons. \label{plot1}}
\end{figure}

Equation (\ref{mbardeen}) cannot be solved analytically, however one can always plot $M=M(r_h)$ and the existence of the extremal
configuration, corresponding to the minimum of the function is given by

\begin{equation}
 \frac{dM}{dr_h}=0\longrightarrow r_{extr.}(Q)= \sqrt{2G_N}Q\ ,\quad M_{extr.}(Q)=\frac{3\sqrt{3}}{4\sqrt{G_N}}\, Q
\label{emin}
\end{equation}

Following Bardeen interpretation one can estimate that for the minimal value of $Q$, i.e.  one electron charge, one finds that

\begin{equation}
 Q_{min}=e \simeq \frac{1}{\sqrt{137}}\longrightarrow r_{extr.}(e) \simeq 0.1\, l_{Pl}\ ,\quad M_{extr.}(e)\simeq 0.06 M_{Pl}
\end{equation}

On the other hand, one expects that there should be no sub-Planckian BHs, thus this model has to be taken as a useful theoretical laboratory,
and not a  phenomenologically viable model of Planckian BHs.\\
In spite of this shortcomings we continue to investigating its thermodynamical characteristics starting with the Hawking temperature
\newpage
\begin{equation}
 T_H =\frac{1}{4\pi}\frac{r_+}{r_+^2 + Q^2 G_N}\left(\, 1 - \frac{r_{extr}^2}{r_+^2}\,\right)
\label{thb}
\end{equation}

Equation (\ref{thb}) shows the well known behavior of BHs admitting an extremal configuration and terminating in a frozen zero temperature 
remnant.\\
Another important thermodynamical characteristic is the BH entropy. Contrary to the usual assumption of the ``universal'' validity
of the celebrated area law, we shall recover it from the First Law, which is given by $dM = T_H dS + \phi_H dQ$

\begin{figure}[h!]
\centerline{\psfig{file=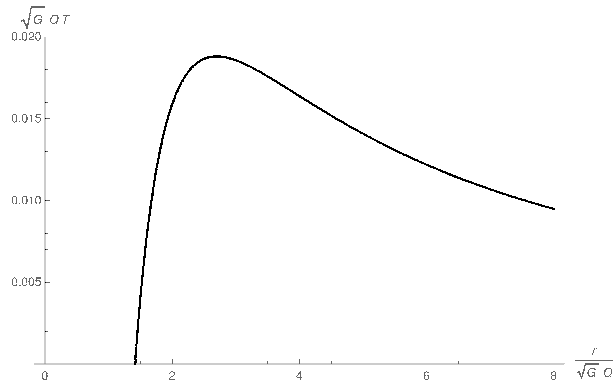,width=8cm}}
\vspace*{8pt}
\caption{Plot of the Hawking temperature (\ref{thb}). \label{plot2}}
\end{figure}

\begin{eqnarray}
 dM &&= \frac{\partial M}{\partial r_+}dr_+ + \frac{\partial M}{\partial Q}dQ\ ,\nonumber\\
    &&= \frac{1}{2G_N r_+^3}\left(\, r_+^2 + Q^2 G_N\,\right)^{1/2} \left(\, r_+^2 - 2Q^2 G_N\,\right)dr_+ + \phi_H dQ\ ,\\
\phi_H &&=\frac{3Q}{2r_+^2}\left(\, r_+^2 + Q^2 G_N\,\right)^{1/2}
\end{eqnarray}

\begin{equation}
 dS= \frac{2\pi}{G_N r_+^2}\left(\, r_+^2 + Q^2 G_N\,\right)^{3/2} dr_+
\end{equation}

and

\begin{equation}
 S= \frac{2\pi}{G_N}\int_{r_{extr.}}^{r_+}\frac{\left(\, r^2 + Q^2 G_N\,\right)^{3/2}}{r^2} dr
\label{ints}
\end{equation}

\begin{eqnarray}
 \int dx \frac{\left(\, x^2 + a^2\,\right)^{3/2}}{x^2} &&= x^2\, \left[\,\frac{3}{2} -\left(\, 1 +\frac{a^2}{x^2}\,\right)^{3/2}\,\right]+
\frac{3a^2}{2} \sinh^{-1}\left(\,\frac{x}{a}\,\right)\ ,\nonumber\\
 &&= x^2\, \left[\,\frac{3}{2} -\left(\, 1 +\frac{a^2}{x^2}\,\right)^{3/2}\,\right]+
\frac{3a^2}{2}\ln\left[\, \frac{x}{a} +\sqrt{1 +\frac{x^2}{a^2} }\,\right]\ ,\\
&&\equiv I(x)
\end{eqnarray}

\begin{equation}
 S = \frac{\pi}{G_N}r_+^2\left[\, \frac{3}{2} -\left(\, 1 +\frac{r^2_{extr.}}{2r_+^2}\,\right)^{3/2}\,\right]
+\frac{3 r^2_{extr.}}{4}\ln\left[\, \frac{\sqrt{2}r_+}{r_{extr.}} + \sqrt{1 + \frac{2r^2_+}{r_{extr}^2}} \,\right]-I(r_{extr.})
\end{equation}

The following  remarks  are in order:
\begin{itemize}
 \item in the limit $Q=0$,  one recovers  the area law
\begin{equation}
 S= \frac{\pi}{G_N}r_+^2=\frac{A_H}{4G_N}
\end{equation}
\item It is usually assumed that only by taking into account $1$-loop gravitational corrections there should be a logarithmic correction
to the area law. Here, we have shown that there are logarithmic corrections already at the semi-classical level, as soon as, one considers
a non point-like source \cite{Nicolini:2012fy,Isi:2013cxa,Nicolini:2012eu}. 
At this point one may wonder what is the matter source leading to (\ref{uno}), (\ref{due}). To answer this question
we out-line the reverse engineering procedure for finding a matter source from a give metric of the form (\ref{uno}).  From the
Einstein equations one gets

\begin{equation}
 m(r)=4\pi \int_0^r dx x^2 \, \rho(x)
\end{equation}

and

\begin{equation}
 \rho(r)= \frac{1}{4\pi r^2} \frac{dm}{dr}
\end{equation}

In case of (\ref{due}) these general formulae lead to

\begin{equation}
 \rho(r)= \frac{3}{4\pi}\frac{M Q^2 G_N}{\left(\, r^2 + Q^2 G_N\,\right)^{5/2}}
\label{rhob}
\end{equation}
{  The plot of $\rho(r)$ is given in Fig.(\ref{ob})}

\begin{figure}[h!]
\centerline{\psfig{file=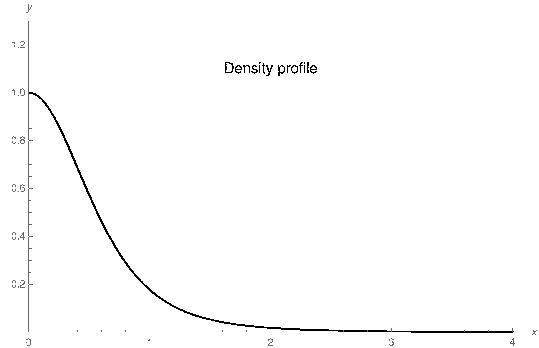,width=7cm}}
\vspace*{8pt}
\caption{Plot of the density (\ref{rhob}), $y\equiv 4\pi Q^3 G^{5/2}/3M$, $x\equiv r/Q\sqrt{G_N}$. \label{ob}}
\end{figure}

\end{itemize}

At this point a brief summary of what we have learned from this toy-model is in order. Although we have used Bardeen model
we give it a different physical interpretation. Instead of advocating some vague non-linear electro-dynamical effects,
which usually lead to unnecessary, but unavoidably, complicated models, we boiled down everything to a neutral BH geometry
sourced by a regular matter distribution (\ref{rhob}), where $Q\sqrt{G_N}$ is simply related to the characteristic size of the
profile.\\
The finiteness of $\rho(0)=3M/4\pi Q^3 G_N^{3/2}$ forbids the presence of an unphysical curvature singularity in $r=0$,  which is
instead replaced by a smooth de Sitter core. Additional important feature of this (and similar 
\cite{Hayward:2005gi,Spallucci:2014kua,Bonanno:2000ep}) 
model is the existence of a lower bound to the BH mass spectrum: the lightest object being ans extremal BH. Heavier objects result in a double-horizon BH
even for neutral, non-spinning object. This is a novel feature, with respect to the Schwartzschild case, shared by regular models to
be discussed in the next sections.

\section{Gaussian BHs}
\label{gbh}

In the previous  Section[\ref{proto}] we have described a model in which  the length cut-off $Q\sqrt{G_N}$ was introduced in the metric 
in order to render it regular. Afterwards, the source of the field, i.e. the energy momentum tensor, was obtained from the Einstein equations
through a procedure called ``reverse-engineering''.\\
However, the textbook procedure is to solve Einstein equations with an assigned energy-momentum tensor.
The simplest physical source of gravity is a single point-like particle. It is clear that a non-vanishing
mass $M$ concentrated in a zero volume leads to an infinite density  and one cannot expect that such a source results in a regular
geometry. However, a point-like mass is only an idealization for a very
a physical object of very small volume. On the other hand, already in  non-relativistic quantum mechanics point-like objects
spread as a consequence of the Uncertainty Principle. The best possible localization for a free quantum particle is given by a Gaussian 
wave-packet whose squared modulus leads to a Gaussian density of the form

\begin{equation}
\rho(r)= \frac{M}{(2\pi)^{3/2} a^3} e^{-r^2/2a^2} \label{g1}
\end{equation}

The corresponding  energy-momentum tensor will be chosen to be the one of an ``anisotropic fluid'': 
$T^{\mu\nu}=diag\left[\, \rho(r)\ , p_r(r)\ , p_\perp(r)\ ,p_\perp(r)\,\right]$. {  This form of $T^{\mu\nu}$ leads to regular versions of
known BH geometries as shown in \cite{Dymnikova:1992ux,Nicolini:2005vd}}.\\
The tangential pressure $p_t$ is determined in terms of $\rho$  from the condition  $D_\mu\, T^{\mu\nu}=0$

\begin{equation}
 p_\perp = -\rho -\frac{r}{2}\partial_r \rho\ .
\end{equation}

$p_(r)$ is let free and must be assigned  in terms $\rho$ in order to fully specify the characteristics of the source through the equation of 
state.
As our source is a \emph{quantum} particle, it is not appropriate  to use ``classical'' equations of state. 
In the previous section we have shown ( equation
(\ref{dslim})), that at short distance, the regular metric approaches the  de Sitter one.  On the other hand, it is known 
that the de Sitter geometry solves the Einstein equations sourced by the vacuum described by the equation of state $\rho= -p= const.$
It is natural  to generalize  this equation  as

\begin{equation}
 \rho(r)=-p_r(r)
\end{equation}

The physical meaning of this assignment is that the outward pressure prevents a gravitational collapse of the matter source to a singular
state, also justifying the choice of a finite width Gaussian profile for $\rho(r)$.
 The  width $a$ is a measure of the quantum particle de-localization, i.e. $\Delta r\simeq a$. \\
Nevertheless, as long as the (quantum) object is of size $a>> l_{Pl}$, gravity will ``see'' it as a \emph{classical}
mass distribution. This important feature, often overlooked in the literature, justify the use of a quantum particle density (\ref{g1})
as the source in the \emph{classical} Einstein equations. If this were not possible then General Relativity should be replaced by a full 
quantum theory of gravity. 
It is also implied  that any ``semi-classical'' description will become less and less reliable as  $a\to l_{Pl}$. 
In the last part of this paper we shall  describe an attempt to develop a full quantum description for a Planckian BH.
\\
By solving the field equations one finds a Schwartzschild-type solution where the mass is quantum mechanically spread 

\begin{eqnarray}
 && ds^2=-f(r)dt^2 +f^{-1}dr^2 +r^2 \left(\, d\theta^2 +\sin^2\theta \, d\phi^2\,\right)\ ,\label{s1}\\
 && f(r)\equiv 1 -\frac{2MG_N}{r}\frac{\gamma\left(\, 3/2\ ; r^2/2a^2\,\right)}{\Gamma(3/2)}\ ,\label{s2}\\
 && \gamma\left(\, 3/2\ ; r^2/2a^2\,\right)\equiv \int_0^{r^2/2a^2}dt t^{1/2} e^{-t} \label{s3}
\end{eqnarray}

At large distance, i.e. for $r>> a$, the incomplete Gamma-function $\gamma\left(\, 3/2\ ; r^2/2a^2\,\right)\to \Gamma(3/2)  $
and we recover the textbook Schwarztschild metric. \\
If we momentarily ``forget'' the physical hypothesis leading to the solution (\ref{s1}),(\ref{s2}),(\ref{s3}), i.e.
 we suspend  the ``quantum'' interpretation of the source and ignore that in General Relativity distances smaller than
the Planck length are physically meaningless,  we can inquire the (classical) space-time short distance
behavior, as well. For $r<< a$:

\begin{equation}
 \gamma\left(\, 3/2\ ; r^2/2a^2\,\right)\approx \frac{1}{3\sqrt{2}}\frac{r^3}{a^3}
\end{equation}

and the line element represents a de Sitter geometry

\begin{equation}
 ds^2 \approx -\left(\, 1 - \frac{\Lambda_{eff}}{3} r^2\,\right) dt^2 +\left(\, 1 - \frac{\Lambda_{eff}}{3} r^2\,\right)^{-1} dr^2 +
r^2 \left(\, d\theta^2 +\sin^2\theta \, d\phi^2\,\right)
\end{equation}

with an \emph{effective cosmological constant}

\begin{equation}
 \Lambda_{eff} \equiv 2\sqrt{\frac{2}{\pi}}\frac{MG_N}{a^3}
\end{equation}

A lengthy calculation of the Kreschmann invariant to prove that there is no curvature singularity in $r=0$ can be skipped as
the de Sitter geometry is \emph{regular} everywhere, including $r=0$.\\
Thus, even  from a purely mathematical point of view, there is no singularity in the geometry   (\ref{s1}),(\ref{s2}),(\ref{s3}).\\
The de Sitter metric describes a non-trivial vacuum geometry where

\begin{equation}
 \rho_\Lambda= -p_\Lambda = \frac{\Lambda}{8\pi G_N}
\end{equation}

In our case 

\begin{equation}
 \rho_\Lambda=\rho(0)= \frac{M}{(2\pi)^{3/2} a^3}
\end{equation}

This result shows that the core of the wave-packet, where the mass density is to a good approximation constant, behaves as a non-trivial
vacuum domain. In this region the negative pressure provides the balancing force stopping the mass shrinking to a singularity.
Recalling our original quantum picture, we can see that the de Sitter core is the gravitational ``translation'' of the uncertainty
principle forbidding the particle to turn into a singularity in $r=0$. Even without taking this picture too literally, it suggests as
quantum effects can eliminate unphysical singularities. \\
An important feature of the metric (\ref{s1}),(\ref{s2}), (\ref{s3}) is that horizons exist only above a minimum mass.  To find this
minimum mass we have to find the zeros, $r=r_h$, of $f(r)$ and its  first derivative:
\begin{eqnarray}
 && f(r_h)=0\longrightarrow \gamma\left(\, 3/2\ ; r^2_h/2a^2\,\right)=\frac{r_h}{2MG_N} \Gamma\left(\, 3/2\,\right)\ ,\\
&& f^\prime(r_h)=0\longrightarrow \gamma\left(\, 3/2\ ; r^2_h/2a^2\,\right)=r_h \gamma^\prime\left(\, 3/2\ ;r^2_h/2a^2 \,\right)
\end{eqnarray}

The first equation identifies the zeros as horizons of the metric, and the second one is the condition for the ADM mass $M=M(r_h)$
is minimal. From the second equation we recover $r_h = r_h(a)$, and replacing in the first equation we find the corresponding value
of $M$.\\
A solution can be found by plotting $M=M(r_h)$ which has the same shape as the curve in Fig.(\ref{plot1}). In the same way, the temperature
is quite similar to the plot in Fig.(\ref{plot2}).\\
So far, we kept the parameter $a$ arbitrary. The natural way to give $a$ a physical meaning is to identify it with the Compton length
of the quantum particle. This identification  is due to the fact that $\lambda$ is a measure of the quantum spread of the particle location.\\
It is interesting to recall that the Compton wave length of a Planck mass particle matches its Schwarzschild
radius and becomes smaller if we further increase the mass(energy).  From the point of view of an external observer the particle has
turned into a BH. This transition has been recently advocated to shield sub-Planckian distances from any experimental
probe, leading to the the UV self-completeness of quantum gravity 
\cite{Dvali:2010bf,Dvali:2010jz,Dvali:2011th,Dvali:2012mx,Dvali:2014ila,Carr:2014mya}. 
It is interesting to see if any of these ideas can be realized in our model.

\begin{figure}[h!]
\centerline{\psfig{file=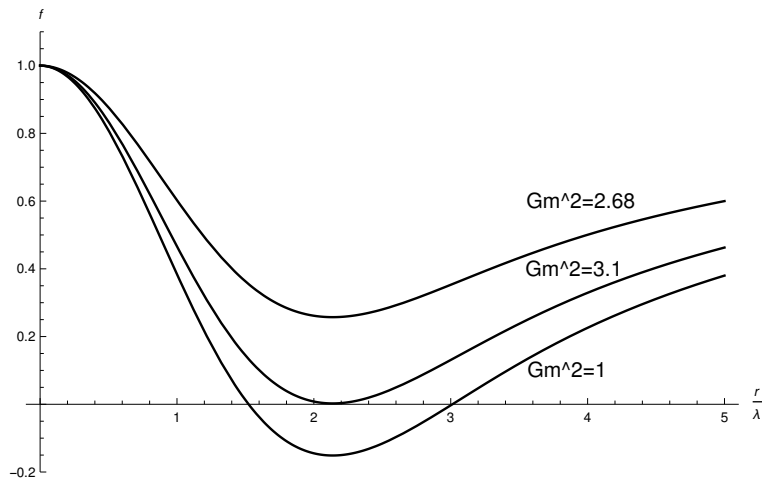,width=10cm}}
\vspace*{8pt}
\caption{Plot of the metric function (\ref{cmetric}). $m^2 < 1.34 M_{Pl}^2$ there are no horizons. For 
$m^2 \ge 1.34 M_{Pl}^2$ the metric describe a BH.\label{cmetrica}}
\end{figure}
\begin{equation}
 a\equiv \lambda =1/m
\end{equation}

Now, equation (\ref{s2}) takes the form
\begin{equation}
 f(r)\equiv 1 -\frac{2mG_N}{r}\frac{\gamma\left(\, 3/2\ ; r^2m^2/2\,\right)}{\Gamma(3/2)} \label{cmetric}
\end{equation}
Equation (\ref{cmetric}) shows a non-linear dependence from the mass $m$ which now appears also in the argument of the incomplete
gamma function. Fig.(\ref{cmetrica}) shows that the metric (\ref{cmetric}) describes both particles and BHs, for different values of $m$ which
is the only free parameter left.  Thus taking $\lambda$ as the width of the Gaussian leads to the qualitative realization of the
UV self-complete quantum gravity program in the following sense:
the particle-BH transition takes place for $m=m_0\simeq 1.6\, M_{Pl}$ 
an extremal BH. Below this mass there are only quantum particles. Above this mass we have double-horizon BHs.

\section{Maximum density}
\label{maxdens}
The discussion in the previous section led to some interesting conclusion that we would like to analyze further. In particular,
the existence of the minimum mass extremal BH sets the upper bound to the possible density of a quantum particle. Let us
estimate this limiting density. One finds

\begin{equation}
 \rho_{max}\equiv \frac{3}{4\pi}\frac{m_0}{r_0^3}\approx 0.73\, \rho_{Pl}
\end{equation}

To a very good approximation, one can assume that no physical object can reach densities above $\rho_{Pl}$. 
Thus, although gravitational collapse is the most efficient compression mechanism in Nature, even in this case 
\emph{ matter  cannot reach densities beyond  $\rho_{Pl}$.} This limit provides the ultimate barrier which prevents 
the formation of any singularity in the space-time fabric. Furthermore, in the Planckian-phase matter building blocks cannot be individually
distinguished anymore because there is no physical probe with wavelength smaller than $l_{Pl}$ to resolve their mutual
distance.  In the words of \cite{Schiller:1996fw}:
''\textit{One thus finds that in a volume of Planck size, it is impossible to say if there
is one particle or none when weighing it or probing it with a beam! In short, vacuum,
i.e. empty space-time can not be distinguished from matter at Planck scales}`` . 
Therefore, the \emph{only} possible equation of state is the one of the de Sitter vacuum, because energy
and pressure cannot be anymore described in terms of individual ''particles``.\\
According with the introductory discussion,   the central density of a particle is at most
$\rho_{Pl}$. A maximally compact version  of the density (\ref{g1}) is given by

\begin{equation}
 \rho\left(\, r\,\right)= \frac{3}{4\pi G_N^2}\exp\left[\,- \pi^{1/3}\left(\, \frac{3M_{Pl}}{4M}\,\right)^{2/3}\frac{r^2}{G_N}\,\right]
\label{dens}
\end{equation}

Using the Gaussian density (\ref{dens}) in the energy-momentum tensor of an anisotropic fluid introduced previously, Einstein equations 
lead to the metric

\begin{eqnarray}
&& ds^2 = -f(r)dt^2 + f^{-1}(r) dr^2 + r^2 d\Omega^2\ ,\\
&& f(r)= 1 -\frac{2m(r)\,G_N}{r}\ ,\label{sol}\\
&& m(r)= \frac{M}{\Gamma(3/2)}\gamma\left[\, \frac{3}{2}\ ; \pi^{1/3}\left(\, \frac{3M_{Pl}}{4M}\,\right)^{2/3}\frac{r^2}{G_N}\,\right]  
 \label{mrad}
\end{eqnarray}
The effective description of quantum effects is still in terms of classical geometry.
However, the memory of the underlying quantum effects is encoded in the non-linear dependence on mass both in the density (\ref{dens})
and the the metric function $f(r)$.  One can verify that the metric (\ref{sol}) shares all the basics features of the solution (\ref{uno}).
For example, we shall explicitly verify  the existence of multiple horizons. In spite of the complicated non-linear mass dependence of the
metric, the existence of horizons can be inferred by plotting $f(r)$ for different values of $M$.
\begin{center}
\begin{figure}[h!]
\includegraphics[height=5cm]{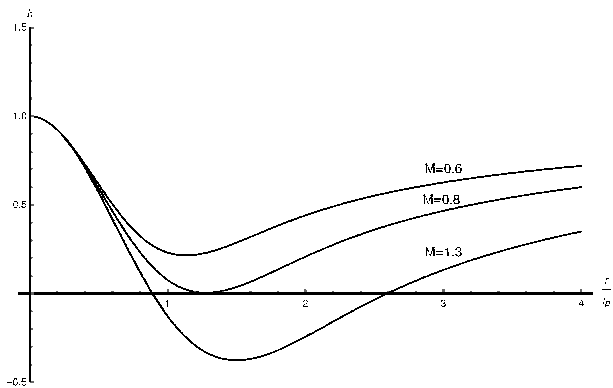}
\caption{Plot of the metric function $h(r)$. For different masses there are two horizons ( $M > 0.8 M_{Pl}$);
one extremal horizon ( $M=0.8 M_{Pl}$); no horizons ( $M < 0.8 M_{Pl}$). }
\label{metrica}
\end{figure}
\end{center}

The plot in Figure (\ref{metrica}) describes the cases  $M\le M_0$ or $M> M_0$, $M_0\simeq 0.8 \,M_{Pl} $:
\begin{itemize}
 \item For $M< M_0$ there are no horizons and space-time is time-like and regular everywhere. 
 \item For $M=M_0$ there is a single, degenerate, horizon of radius $r_0 \approx 1.29\, l_{Pl}$ and $h(r)$ describes an
 \emph{extremal} BH with the smallest mass.
  \item For $M> M_0$ there are two, non-degenerate,  horizon and the solution describes a regular BH.
\end{itemize}

The above discussion follows the same pattern   as the one in the Section[\ref{proto}], and also agrees with the results obtained in
\cite{Nicolini:2005vd}. 
It is safe to conclude that the described behavior of regular BHs is a model independent feature.
In particular, the existence of a minimum mass $M_0$, 
gives the possibility to provide a quantitative formulation of the \emph{hoop conjecture} \cite{thorne,Mureika:2011hg,Casadio:2013uga} 
for non-homogeneous  masses. For this purpose, let us plot both the radial mass distribution $m(r)$ and the density $\rho(r)$

\begin{center}
\begin{figure}[h!]
\includegraphics[height=5cm]{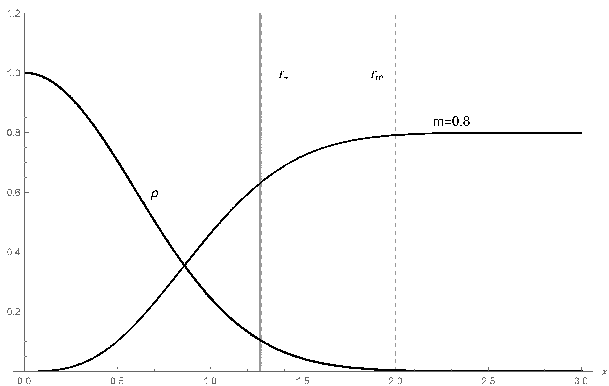}
\caption{ Plot of the density, dashed line, and radial mass distribution, continuous line. The vertical dotted line
marks the position of the extremal horizon. The intersection between these two curves gives the fraction of mass within the horizon. Only
a small fraction surrounds the horizon from the outside.}
\label{metrica2}
\end{figure}
\end{center}

Figure (\ref{metrica2}) shows that, in order for a  BH of mass $M= M_0$ to be formed, it is enough to have, approximately,  
$72\%$ of the total mass $M$  within a sphere of radius $r_0 \approx 1.29\, l_{Pl}$. Therefore, in this picture the extremal BH is 
as close as possible to the classical ''hidden star`` in the sense that its interior is completely filled with matter, 
though non-uniformly distributed. \\ 
For $M> M_0$ there are two horizons $r_\pm$, $r_- < r_+$. The inner horizon is deep into the Planck density matter core surrounding
the origin. Thus, it is physically unreachable by any probe. On the other hand,  $r_+$ is at the border of the BH ''~\emph{atmosphere}~``,
where the matter density is close to zero. In other words, non-extremal BHs are almost empty as in the relativistic formulation, though
preserving a central, non-singular, massive core.\\



In the present case, the  Hawking temperature  is given by

\begin{equation}
 T=\frac{1}{4\pi r_+}\left[\, 1 -\frac{r_+^3}{\sqrt{2}a^3}\frac{ e^{- r_+^2/2a^2}}{\gamma\left(\,\frac{3}{2}\ ;  r^2_+/2a^2 \,\right) }\,\right]
\ ,\quad a^2\equiv \left(\, \frac{2}{\pi}\,\right)^{1/3} \left(\, \frac{M}{3M_{Pl}}\,\right)^{2/3} l^2_{Pl}
\label{tgen}
\end{equation}

which is  an implicit function of $r_+$, since it is also dependent on $M$:

\begin{equation}
 M=\frac{r_+}{2G_N\Gamma(3/2)}\gamma\left[\, \frac{3}{2}\ ;\left(\, \frac{3M_{Pl}}{4M}\,\right)^{2/3} \frac{r^2_+}{G_N}\,\right]
\label{emme}
\end{equation}

Therefore, it is not possible to plot (\ref{tgen}) without some kind of 
approximation. A fairly good result for  
$M$ is obtained by solving iteratively equation (\ref{emme}). At the first oder one finds:

\begin{equation}
 \frac{1}{2a^2} \approx \left(\, \frac{3\pi^{1/2}}{8G_N r_+}\,\right)^{2/3}
\label{approx}
\end{equation}

\begin{center}
\begin{figure}[h!]
\includegraphics[height=5cm]{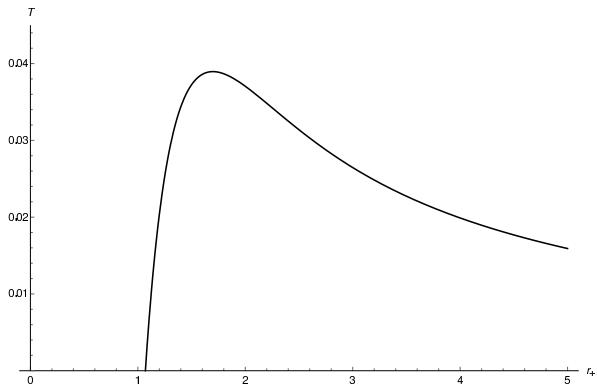}
\caption{Plot of the  Hawking temperature (\ref{approxT}), in Planck units, for (approximated) $M\simeq r_+/2G_N$.}
\label{HT}
\end{figure}
\end{center}

and the approximated version of the temperature, shown in Fig.(\ref{HT}) is

\begin{equation}
 T\simeq\frac{1}{4\pi r_+}\left[\, 1 -\frac{3\pi^{1/2}r_+^2}{4G_N}\, 
\frac{e^{-\left(\, 3\pi^{1/2}r_+^2/8G_N\,\right)^{2/3}}}{\gamma\left(\,3/2\ ; \left(\, 3\pi^{1/2}r_+^2/ 8G_N\,\right)^{2/3}  \,\right) }
\,\right]\label{approxT}
\end{equation}

To estimate the validity of the approximation we compare the value of the extremal radius, at which $T=0$, in Fig.(\ref{HT}) and
compare it to the true value of the extremal BH radius as obtained in Fig.(\ref{metrica}).
The discrepancy between the two radii is only about  $2\%$  .\\
In general terms, the plot in Fig.(\ref{HT}) reproduces all the interesting features typical  for regular, Gaussian, BHs. In particular,
the temperature  is always finite, and  vanishes for the extremal configuration.  Therefore, even though they approach Schwarzschild form 
for $r_+ >> l_{Pl}$, they show a very different behavior at small distances.
In this way, all the ''anomalies`` of the final stage of evaporation are cured in a semi-classical framework which, however, encodes the
fundamental information about the finiteness of the Planck density, as it follows from quantum uncertainties at this energy scale.\\

Returning back to the comparison between classical and relativistic models of BHs, already discussed in relation to Fig.(\ref{metrica2}),
the approximation (\ref{approx}) leads to

\begin{equation}
 \frac{r_+}{a}\propto \left(\, \frac{M}{M_{Pl}}\, \right)^{2/3}
\label{ratio}
\end{equation}
 
Equation (\ref{ratio}), again, confirms that, for non-extremal BH with $M> M_{Pl}$, the horizon radius is much larger than the with of 
the matter distribution. Thus, non extremal BHs are ''~almost empty~``, as the major part of their  mass is enclosed in an inner sphere 
much smaller than the horizon itself.\\
The main idea of this section is that any gravitationally collapsing object of \textit{arbitrary} mass should  never exceed the Planck density
at its core.  This gives a universal picture for the ultimate stage of matter compression by self-gravity.  
An immediate consequence is the absence of curvature singularity as the final stage of a gravitational collapse.  
We have already shown in a series of previous papers how to avoid  singularities  by implementing in the Einstein equations a quantum gravity
induced ''~minimal length~``. In the absence of a general consensus about what a quantum theory of gravity should be,
one could question the physical origin of a minimal length. We have shown in this section that introducing an ''~ad hoc~``
 length parameter  can be avoided  by developing an alternative self-consistent,  physically meaningful, model of regular BH. \\
The present description has led to a compromising picture between the classical model of super-dense black stars and the
relativistic view of ''~empty~`` BHs with mass concentrated in a singularity. Removal of the curvature singularity results
in a partial fullness of the BH interior. Among all relativistic BHs, the extremal one is the closest to the classical hidden stars 
 in the sense that its interior is full of non-uniformly distributed matter.\\
In other words, the ''~emptiness~`` of the interior depends on
mass $M$, i.e. for $M>> M_{Pl}$ interior is almost, but not completely, empty, while for $ M \simeq M_{Pl}$ the interior is full.\\
To be fair, there is a small fraction of mass outside of the horizon, which is due to the Gaussian shape of the source. {  This tiny
tail, does not prevent BH formation, but may also provide a possible resolution for the  information paradox.}
In fact, the horizon \emph{remains in contact} with the interior mass and the whole information it encodes. 
Furthermore, this model also offered a quantitative formulation of the
''~hoop conjecture~``. We have shown that it is sufficient to have, approximately, $72\%$ of the total distributed mass 
inside its own gravitational  radius for the BH to appear.

\section{The ``breathing'' horizon: a classical particle-like model }
\label{breath}
 In previous sections,  we have described various models of semi-classical BHs which however contained some quantum in-put.
 This description relied on the quantum improved matter source in the \emph{classical} Einstein equations. The final result is a
 geometrical description of BHs in terms of a quantum improved line element.  It is widely accepted, however, that at the Planck
 scale General Relativity is inadequate description of gravity.  A genuinely new quantum formulation is needed. So far, the most promising
candidate for quantum gravity is (Super)String Theory,  since it naturally incorporates the graviton in the string spectrum. However, even in
the case of \emph{stringy} the description boils down to a, more or less complicated, classical metrics \cite{libro}.
 Thus, we are again back to the beginning!\\
One of the expectations in future LHC experiments is the appearance of signals indicating the presence
            of Planck scale micro BHs, at least, as virtual intermediate states 
            \cite{Calmet:2008dg,Calmet:2012cn,Calmet:2012fv,Belyaev:2014ljc,Calmet:2014uaa,Calmet:2015fua}. 
            In other words they are supposed to be just another  structure in the elementary particle zoo. 
            From this point of view, it is hardly arguable that these quantum gravitational excitations can somehow defy 
            the laws of quantum mechanics and instead
            be described on the same geometrical terms as their cosmic cousins of a million, or so, solar masses.
            Oddly enough, so far this is the dominant point of view. Occasionally, in the distant past a few
            dissonant ideas have been put forward and largely ignored \cite{Markov:1967lha,Markov:1972sc,Spallucci:1977wc}. 
            Nevertheless, very recently the same line of
            thinking in terms of non-geometric, and purely quantum mechanical description,  has gained ground as  alternative view to 
            the standard geometrical approach. In order to be completely clear,in this approach  one is not thinking in terms of a 
            quantum version of Einstein General Relativity, but rather in terms of a purely particle-like quantum mechanical formulation. 
            The classical horizon, as a smooth boundary surface, is expected to emerge in a suitable classical limit of this quantum picture. 
            Thus, we shall introduce a particle-like model of micro BHs, where the only link with the classical geometric 
            description is through the linear relation between its size and total mass-energy, i.e. horizon equation.\\

             In the absence of any tractable quantum gravity equation to start with, we shall develop a suitable quantization procedure 
             starting from a classical, particle-like, model of the horizon itself. What should such a classical model should be based on?\\
             Certainly, not on the dynamics of a classical BH, described as a single particle subject to external forces.  
              The main obstacle for quantizing  a classical BH is that its horizon is a geometrical 
              surface without internal dynamics. Thus, the canonical quantization of the BH has no classical counterpart to start with. \\
              Therefore, as a first step, the static horizon has to be given a proper classical dynamics. In other words, it will
              be assigned its own kinetic energy and will evolve in time.\\
              In the case of spherically symmetric BH the problem reduces to the single, radial coordinate which is allowed
              to ``~\emph{breath}~'', achieving maximum ``~lung capacity~'' corresponding to the classical  Schwarzschild 
              radius $r_h= 2M G_N$. \\
                        
               The quantization of any  mechanical system starts from a classical Hamiltonian encoding its motion.
On the other hand, a classical BH is defined as a particular solution of the Einstein equations. We give up such a starting
point in favor of a particle-like formulation translating in a mechanical language.  
 In the simplest case of a Schwarzschild BH the particle-like Hamiltonian will be constructed
taking into account the following features:
\begin{enumerate}
             \item BHs are intrinsically \emph{generally relativistic} objects, in the sense of strong gravitational fields.
              Thus, the equivalent particle model should start with 
               a relativistic-like dispersion relation for energy and momentum, rather than a Newtonian one;
             \item the particle model must share the same spherical symmetry  and the classical motion will
                   be described in terms of a single radial  degree of freedom $r_h$;
             \item  the ``~mass~'' associated to the horizon  is the ADM $M_{ADM}=M(r_h)$ which will be identified with 
                    the ``~particle~'' mass. Therefore, the main
                    distinction between an ``~ordinary~'' quantum particle and a QBH is:
                    i) the linear extension of the particle, characterized by its Compton wavelength, decreases with the mass;
                    ii) the linear extension of a QBH, characterized by its horizon radius, increases with its mass.\\
             \item  In our classical BH particle-like model, the horizon equation  turns into the  equation
                    for the turning points of a particle, with total energy $E$, subject to the potential
                    
            \begin{equation}
            V(r_h) =\frac{r_h}{2\, G_N}\ ,\qquad r_h\ge 0 \label{poth}
             \end{equation}
            This is the usual harmonic potential, though restricted to the positive semi-axis $r_h\ge 0$. The curvature singularity
            of the Schwarzschild BH is mimicked by the perfectly reflecting wall in $r_h=0$. Thus, the motion of the particle is restricted
            between the origin and a maximum elongation.
            
                 \emph{These prescriptions allow to map the geometric problem of finding the horizon(s), in a given metric, into the
                 dynamical problem of determining the turning points, for the bounded motion, of a classical relativistic particle.}
                \end{enumerate}
            
         The above requirements lead to the following relativistic Hamiltonian

\begin{equation}
 H\left(\, r_h\ , p\,\right)\equiv \sqrt{p^2_h + M^2(r_h)}=\sqrt{p^2_h +\frac{r_h^2}{4G_N^2} }
\label{hh}
\end{equation}

where, $p$ is the canonical momentum conjugated to the horizon radial coordinate $r_h$. Before solving the equation of motion,
it is worth to comment on the harmonic term in the square root:\\
\begin{itemize}
 \item it is \emph{not} an ad hoc choice, but it  \emph{follows} from the horizon equation. In other words, the potential
is self-consistently generated by the BH itself.
\item The specific form chosen in (\ref{poth}) is harmonic in the Schwarzschild case, and is uniquely determined by the type of the BH
       considered.
      In fact, in case of a Reissner-Nordstrom BH  is not a simple harmonic term,  but also has
      an-harmonic corrections \cite{Spallucci:2016qrv}:
 
\begin{equation}
V_{RN}(r_h)=  \frac{r_h^2}{4\,G^2_N}\left(1+\frac{Q^2G_N}{r_h^2}\right)^2
 \end{equation}
     
\end{itemize}
 
For any conservative system  the Hamiltonian is a  constant of motion:

\begin{equation}
 H=E \label{etot}
\end{equation}

Using the Hamilton equation

\begin{equation}
 \frac{\partial H}{\partial r_h}=-\dot{p}=\frac{r_h}{4G_N^2 E}
\end{equation}

together with 

\begin{equation}
 p_h=\sqrt{E^2 - \frac{r_h^2}{4G_N^2}} 
\end{equation}

 leads to the equation of motion 

\begin{equation}
 \dot{r_h}^2 =  1 - \frac{r_h^2}{4G_N^2E^2}
\label{sette}
\end{equation}

Setting the initial condition as:

\begin{equation}
 r_h\left(\, t=0\,\right)= 0\ .
\end{equation}

the solution of equation (\ref{sette}) is given by

\begin{eqnarray}
&& r_h\left(\, t\, \right)= 2G_N E \, \sin\left(\, \omega t\,\right) \ge 0\\ 
&& \omega\equiv \frac{2\pi}{T}= \frac{1}{2G_N E}
\end{eqnarray}

The oscillation starting from the origin  reaches the maximum elongation at the Schwarschild radius
$r_h=2G_N E$ after half a period $t=T/2$. 
 \\
In order to be able to confront  classical and quantum results, to be obtained in the next Section, we shall calculate the classical 
mean values for $r_h$ and $r_h^2$ defined as time averages over one quarter of a period  

\begin{equation}
\widehat{r_h}\equiv \frac{1}{\pi G_N E}\int_0^{\pi G_N E} dt\, r(t) =  \frac{4}{\pi} G_N E   =\frac{2}{\pi} r_h \label{rcl}
\end{equation}

\begin{equation}
 \widehat{r_h}^2\equiv \frac{1}{\pi G_N E}\int_0^{\pi G_N E} dt \, r^2(t)=2G_N^2 E^2 = \frac{1}{2}r_h^2 \label{r2cl}
\end{equation}

We see that $\widehat{r_h} > r_h/2$ contrary to what one would  naively expect. The physical reason is that the particle
spends more time close to $r_h$ where the approaching speed tends to zero.\\
We stress that the model introduced in this section, does not describe a 
classical BH solution of  Einstein equations. This is not a contradiction because our classical model is not
meant to describe a geometric BH, but it is only a starting point towards the quantum formulation of a Planckian, particle-like, BH. 
It has, however, something in common with a Schwarzschild BH, i.e. the maximal elongation
is equal to the horizon radius. 

\section{Quantum horizon wave equation}
\label{qh}

Equation (\ref{etot}) is the starting point for the quantization of the system. Since we were working in a relativistic
framework already at the classical level, the corresponding quantum equation will be of relativistic type as well.
Applying the canonical quantization procedure

\begin{equation}
 p_h \longrightarrow i \frac{d}{dr_h}\ , \qquad (\hbar \equiv 1)
\end{equation}

we find the quantum analogue of the classical (\ref{etot})
 
\begin{equation}
  \frac{1}{r_h^2}\frac{d}{dr_h}\left(\, r_h^2 \frac{d\psi}{dr_h}\,\right) 
+ \left(\, E^2 - \frac{r_h^2}{4G_N^2}\,\right) \psi =0 \label{weq}
\end{equation}

where, the horizon wave function $\psi$  is normalized as:

\begin{equation}
 4\pi \int_0^\infty dr_h r_h^2 \psi^\ast \psi =1 
\end{equation}

 It would be, in principle, possible to allow  quantum
fluctuations with non-vanishing angular momentum, we limit
ourselves, in this paper, to the simplest possible case of ``~s-wave'' states only. 
The general, more complicated,  model is presented in \cite{Spallucci:2016qrv}.\\ 
At this point, several comments are in order. 
\begin{itemize}
\item To avoid confusion, we remark that equation (\ref{weq}) is not written in a Schwarzschild background geometry,
because we are not dealing with quantum field theory problem   in a classical Schwarzschild background.
Rather, the ``~particle~'' \emph{is} the quantum horizon itself. 
\item
Additional reason justifying the form of the wave equation (\ref{weq}) is the 
Holographic Principle \cite{Susskind:1994vu,Susskind:1998dq,Hooft:1999bw} implying that the
dynamics of the quantum BH must be described in terms of a wave-function for the horizon only. No reference 
to any bulk geometry is allowed.
\item
Finally, quantization naturally leads to a ''\emph{fuzzy}`` horizon which cannot be meaningfully described in terms of a classical
smooth surface.  The very distinction between the ''interior`` and ''exterior`` of the BH is no more  significant than
the distinction between the interior/exterior of a quantum wave-packet. Therefore, a Planckian BH is to be seen just as another
quantum particle, but with a particular relation between its mass and linear extension.
\end{itemize}


The solution of the equation (\ref{weq}) is:

\begin{equation}
 \psi_n \left(\, r_h^2/2G_N\,\right) = N_n\, e^{-r_h^2/4G_N }\, L_n^{1/2}\left(\, r_h^2/2G_N\,\right)
\end{equation}

where the normalization constant is given by:

\begin{equation}
 N_n=\frac{\sqrt{n!}}{ 2\sqrt{\sqrt{2}\pi G_N^{3/2} \Gamma\left(\, n+3/2\,\right)}} \label{norm2}
\end{equation}

The corresponding quantum BH mass spectrum \cite{Spallucci:2015jea,Spallucci:2014kua} is:

\begin{equation}
 E_n^2=\frac{1}{G_N}\left(\, 2n + \frac{3}{2}\,\right) \ ,\qquad n=0\ ,1\ ,2\ ,\dots
\label{spectrum}
\end{equation}

First thing to remark is the existence of a \emph{ground state} energy, or zero-point energy, near the Planck mass:

\begin{equation}
 E_{n=0}=1.22\times  M_{Pl} \label{vacuum}
\end{equation}

Contrary to the semi-classical description where the mass can be arbitrary small, we find that in a genuine quantum description
the mass spectrum is bounded from below by $E_{n=0}$. In this model the quantization  solves the problem of
the ultimate stage of any process involving emission or absorption of energy.  Neither ''naked singularity`` nor empty Minkowski
space-time are allowed as final stage of the BH decay. The standard thermodynamical picture looses its meaning since
we are in a true quantum regime.\\
The excited states are equidistant much like in the case of an harmonic oscillator.\\
Having acquired the notion that Plankian BHs are quite different objects from their classical ``cousins'', we would like to address
the question of how to consistently connect Planckian and semi-classical BHs. As usual, one assumes that the quantum
system approaches the semi-classical one in the ``large-n'' limit in which the energy spectrum becomes continuous.
Let us first consider the radial probability density  defined as $ p(r_h)\equiv 4\pi r_h^2 \vert \psi \vert ^2 $:

\begin{eqnarray}
 && p_n(x)=\frac{2\, n!}{\Gamma\left(\, n +3/2\,\right)}\, x^2\, e^{-x^2}\, \left(\, L_n^{1/2}\left(\,x^2\,\right)\,\right)^2\ ,\\
 && x\equiv r/\sqrt{2G_N}
\end{eqnarray}

\begin{figure}[pb]
\centerline{\psfig{file=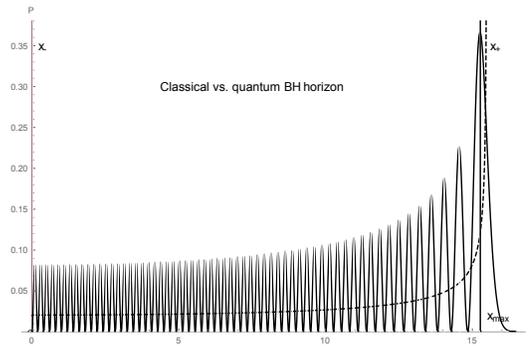,width=7cm}}
\vspace*{8pt}
\caption{ Plot of the function $p_{n=60}(x)$,  (continuous line) vs classical probability (dashed line). For large $n$ the
position of the  peak approaches the classical Schwarzschild radius $r_h= 2E G_N$.. \label{massimi}}
\end{figure}


The local maximum points in figure(\ref{massimi}) represent the most probable sizes of the Planckian BH.  We remark that:\\
\begin{itemize}
 \item there exist an absolute maximum for any $n$. In the ``\emph{classical}'' limit $n\to \infty$, the absolute maximum
       approaches the classical Schwarzschild radius $r_h = 2G_N E$.
 \item Quantum fluctuations allow larger radii but the probability is exponentially suppressed, as shown by the 
       vanishing tail penetrating the classically forbidden region. Furthermore, the penetration depth is quickly
       decreasing as $n$ becomes larger and larger, indeed this the classical limit where quantum fluctuations
       vanish and the most probable value of $r_h$ freezes at the classical value.
\end{itemize}

These maximum points  are solutions of the equation

\begin{equation}
\left(\, 1 -x^2 +4n\,\right)\, L_n^{1/2}\left(\, x^2\,\right)-2\left(\, 2n +1/2\,\right)\, 
L_{n-1}^{1/2}\left(\, x^2\,\right)=0 \label{maxima}
\end{equation}

Equation (\ref{maxima}) cannot be solved analytically , but its large-$n$ limit can be evaluated as follows.
First, perform the division $L_{n}^{1/2}/L_{n-1}^{1/2}$, and then write

\begin{equation}
 L_n^{1/2}\left(\,x^2\,\right)=
P_2\left(\, x^2\,\right)\, L_{n-1}^{1/2}\left(\, x^2\,\right)+R_{n-2}\left(\, x^2\,\right)\label{ratio2}
\end{equation}

where, 

\begin{eqnarray}
&& P_2 =\frac{a_n}{b_{n-1}}\left(\, x^2-2n+1/2\, \right)\\
&& R_{n-2}=c_{n-2} x^{2n-4}+\dots\ , \nonumber\\
&&\>\>\>\>\> \>\>\>\>\>\>=-\left(\, n-1\,\right)\left(\, n-1/2\,\right)\, a_n\,x^{2n-4}+\cdots
\end{eqnarray}

By inserting equation (\ref{ratio2}) in equation (\ref{maxima}) and by keeping terms up order $x^{2n-2}$, the equation
for maximum points turns into

\begin{eqnarray}
&& \left[\,x^2-2n-1\,\right]\,\left[\, x^2-2n+1/2\,\right]+\left[\, 2n+1\,\right]\frac{b_{n-1}}{a_n}=\nonumber\\
 &&  =(n-1)(n-1/2)
\label{max1}
\end{eqnarray}

where the coefficients  are given by

\begin{eqnarray}
&& a_n=\frac{(-1)^n}{n!}\ ,\\
&& b_{n-1}=\frac{(-1)^{n-1}}{(n-1)!}
\end{eqnarray}

Equation (\ref{max1}), for large $n$ reduces to
 \begin{eqnarray}
&& 3n^2= \left(\,x^2-2n\,\right)^2 \nonumber\\
&& x^2= 2n+\sqrt{3} n=3.73n\nonumber
\end{eqnarray}

On the other hand, the classical radius of the horizon is obtained as

\begin{equation}
  \frac{r_h^2}{2G_N}= 2G_NE^2  = 4n 
 \end{equation}
 
which leads to

\begin{equation}
 x^2_+=4n
\end{equation}
 
Thus, we find that most probable value of $r_h$ approaches the Schwarzschild radius  for $E>> M_p$, 
restoring the BH (semi)classical picture.

\section{Conclusions}
\label{end}

In this review paper we have presented a sequence of ideas on regular semi-classical/quantum BHs starting from
early attempts of an ad hoc regularization to the genuine quantum BH in a non-geometrical framework. \\
The semi-classical description is still in terms of classical metrics, but with important quantum in-put. We have 
introduced the energy momentum tensor of a matter source spatially distributed  in a Gaussian way to take into account 
the quantum spread. The state equation characterizing the source violates the weak energy condition, which explains the 
regularity of the solutions. Further justification of the choice $\rho(r)=-p_r(r)$ is motivated by its short distance behavior
reproducing the equation of state of the quantum vacuum. \\
By letting free the width of the Gaussian distribution  one gets a two-parameter dependent, quantum improved, geometry 
with respect to the corresponding source-free solutions. \\
Consistency with the quantum interpretation of the width of the matter distribution, strongly suggests the identification
with the Compton wavelength of the quantum particle sourcing the field. This has led to a slightly more complicated one
parameter (mass) dependent geometry maintaining all the nice features of the Gaussian regular solution. It has further
led to the conclusion that there exists a \emph{finite} ultimate  density of the gravitationally collapsed matter, which
we identified with the Planck density.\\
These semi-classical models are good description of microscopic BHs as long as their mass is much larger than the Planck mass.
As this limit is approached, a true quantum description of BHs is necessary. For this regime we have proposed a novel \emph{non-geometric},
particle-like, description for Planckian BHs.\\
Our construction is focused on the ``~horizon wave function~'', as it can be expected from the Holographic Principle.
The same term was introduced  
by Casadio and coworkers \cite{Casadio:2013hja,Casadio:2013iqc,Calmet:2015pea}  to define the probability amplitude
for a point particle to be inside its own Schwartzschild horizon. However, apart from the name
there are substantial differences, in our approach, worth pointing out to avoid confusion and possible misinterpretations.\\
In the former case, a particle of mass $m$ end energy $E$ is identified with a BH whenever it fits inside its own
Schwarzschild radius, given by the \emph{classical} relation $R_H= 2 G_N E$. Thus, these authors start with a 
 a spherical Gaussian wave packet of assigned width, and study only the behavior of the matter
source itself. \\
On the contrary, in our case, there is no reference to any matter source. Rather,
we start from the Holographic Principle, where  the single degree of freedom is represented by the  
BH surface. Therefore, our  equation (\ref{weq}) does not describe a  particle, but it is rather the quantum equation for the horizon itself.
In this way, the dynamical degree of freedom is the  quantum fluctuating BH boundary. When the amplitude of this oscillations becomes
comparable with  the Planck length,  the 
very distinction between  BH interior and exterior is blurred away. Consequently  the Planckian BH is completely different from its
classical counterpart.\\ 
The dynamical QBH model, described in this paper, can also be linked to the  non-geometrical approach by Dvali and co-workers 
\cite{Dvali:2011aa,Dvali:2015aja,Dvali:2012mx,Dvali:2012en}, in which a quantum BH is described as an $N$
graviton BEC condensate. We tentatively identify their characteristic occupational number $N$ with our principal quantum number $n$ in
(\ref{spectrum}) as 
\begin{equation}
 N \equiv 2n
\end{equation}
Nevertheless,  our approach is different from the one of the quoted authors as we do not assume any ad hoc potential 
potential trapping the gravitons. On the contrary, we \emph{derive} in a 
self-consistent way the potential from the classical(geometric)
horizon equation. In the simplest case of a neutral, non-spinning, BH the potential turns out to be harmonic, but in a more general, e.g. 
charged case, non-harmonic terms will appear as well. \\
Furthermore, it is shown in
figure (\ref{massimi}) that a smooth classical boundary surface emerges only in the limit $E>> M_{Pl}$, or $n>> 1$.
In our picture the, so-called, ``classicalization'' , i.e. the transition from a quantum particle to a Planckian  BH, is 
not an abrupt event occurring, as soon as, the Planck mass scale is reached (from below). Rather, there is an intermediate, genuinely
quantum BH phase, which is characterized by the absence of a classical geometrical event horizon. Our quantum gravitational
excitations are  reminiscent of the `` black hole precursors'', appearing  
as complex poles in the ``~dressed~'' graviton propagator description in \cite{Calmet:2014gya}. \\
Last but not the least, a connection to a classical geometrical
 Schwarzschild BH emerges, far above the Planck scale, even if the BH still remains a microscopic object. The new,intermediate , genuinely 
quantum phase, for $E> M_{Pl}$,$r_h > l_{Pl}$,  is a novel feature of classicalization in our model.\\
The model in Section[\ref{qh}] realizes, in a surprisingly simple manner, the growing belief
that Planckian scale are profoundly different  from classical, gravitationally collapsed, objects.
This  behavior turns  ''~dreadful~``  classical BHs into harmless  quantum ``~black~'' particles.\\

\end{document}